\documentclass[twocolumn,showpacs,preprintnumbers,amsmath,amssymb]{revtex4}

\usepackage{graphicx}
\usepackage{dcolumn}
\usepackage{bm}

\begin{document}

\title{Mechanisms of dissipation in wet granular matter}

\author{A. Fingerle}
\email{axel.fingerle@ds.mpg.de}
\author{S. Herminghaus}
\email{stephan.herminghaus@ds.mpg.de}
\affiliation{Max-Planck-Institute for Dynamics and
Self-Organization, Bunsenstr. 10, 37073 G\"ottingen, Germany}

\date{\today}

\begin{abstract}

The impact dynamics between wet surfaces, which dominates the
mechanical properties of wet granular matter, is studied both
experimentally and theoretically. It is shown that the hysteretic
formation and rupture of liquid capillary bridges between adjacent
grains accounts reasonably well for most relevant cases of wet
granular matter. The various dissipation mechanisms are discussed
with particular emphasis on their relevance. Variations of the
rupture energy loss with the impact energy are quantified and
discussed.

\end{abstract}


\maketitle

The interest in granular materials among the soft-matter community
has been strongly increasing in recent years. It has been widely
recognized that many concepts which are well established for
colloidal systems and glasses apply as well to granular systems
\cite{Goldhirsch96,Coniglio04,BrillPosch,BehringerSperl06,Majmudar07},
and vice versa \cite{Ruiz02,AsteConiglio04,Ruiz05,Keys07}.
Furthermore, both colloidal and granular matter play a certain role
as models for other systems which are too complex to be tractable.
In particular, granular systems are of great interest in the context
of dynamical systems far from thermal equilibrium
\cite{FeitosaMenon04,Aumaitre04,Herminghaus05}.

The term 'soft matter' applies particularly well to {\it wet}
granulates, which can be shaped to stable structures
\cite{SandWorldDe}, but yield to rather small shear stress
\cite{Hornbaker97,Mikami98,Bocquet04,Scheel04,Richefeu06}. This
plasticity stands out against dry granulates, such as the sand in an
hour glass, which runs through the orifice like a fluid
\cite{Duran00}. The reason for this difference is that in the wet
system, small liquid capillary bridges form between adjacent grains,
exerting an attractive force upon them by means of the surface
tension of the liquid \cite{Thornton98,Simons00}. It is clear that
these only form when the liquid wets the material the grains consist
of, which is well fulfilled for most sands. When a wet granulate is
being sheared, or otherwise mechanically agitated, the repeated
formation and rupture of the many liquid objects inside gives rise
to considerable dissipation, which is then experienced as a
noticeable resistance to the external drive imposed on the material.
In order to understand the mechanical properties of wet granular
matter, it is thus indispensable to understand the dissipation
processes connected to the liquid capillary bridges in detail.

As it is well known, there is some intrinsic dissipation also in dry
granular matter, which is responsible for the fact that even the
perfectly dry granulate in the hour-glass behaves distinctly
different from a regular fluid. Some fraction of the kinetic energy
of the grains is transferred at each impact to the microscopic
degrees of freedom on atomic scale. The 'heat bath' represented by
the random center-of-mass motion of the grains, which easily
corresponds to Giga- or even Tera-Kelvins when converted to
temperature using Boltzmann's constant, is thus intimately coupled
to the room-temperature heat bath of the atoms. Some of the most
striking features of granular motion owe to this intrinsic
non-equilibrium character. One usually quantifies these effects by
means of the so-called restitution coefficient, $\varepsilon$, which
is defined as the ratio of the momenta before and after the impact,
$p_f = \varepsilon \ p_i$. It is found that this ratio is fairly
independent of the initial kinetic energy of the grains in a wide
range, although for large energies it tends to be somewhat smaller
\cite{BrillPosch}. The most important aspect of this dissipation
mechanism is that the energy lost in the impacts, $\Delta E_{dry}$,
scales with the impact energy, $E$, according to
\begin{equation}
\Delta E_{dry} = (1-\varepsilon^2) E
\end{equation}
where $\varepsilon$ is approximately constant. There is thus no
specific energy scale set by this process.

\begin{figure}
\includegraphics[width = 6.5cm]{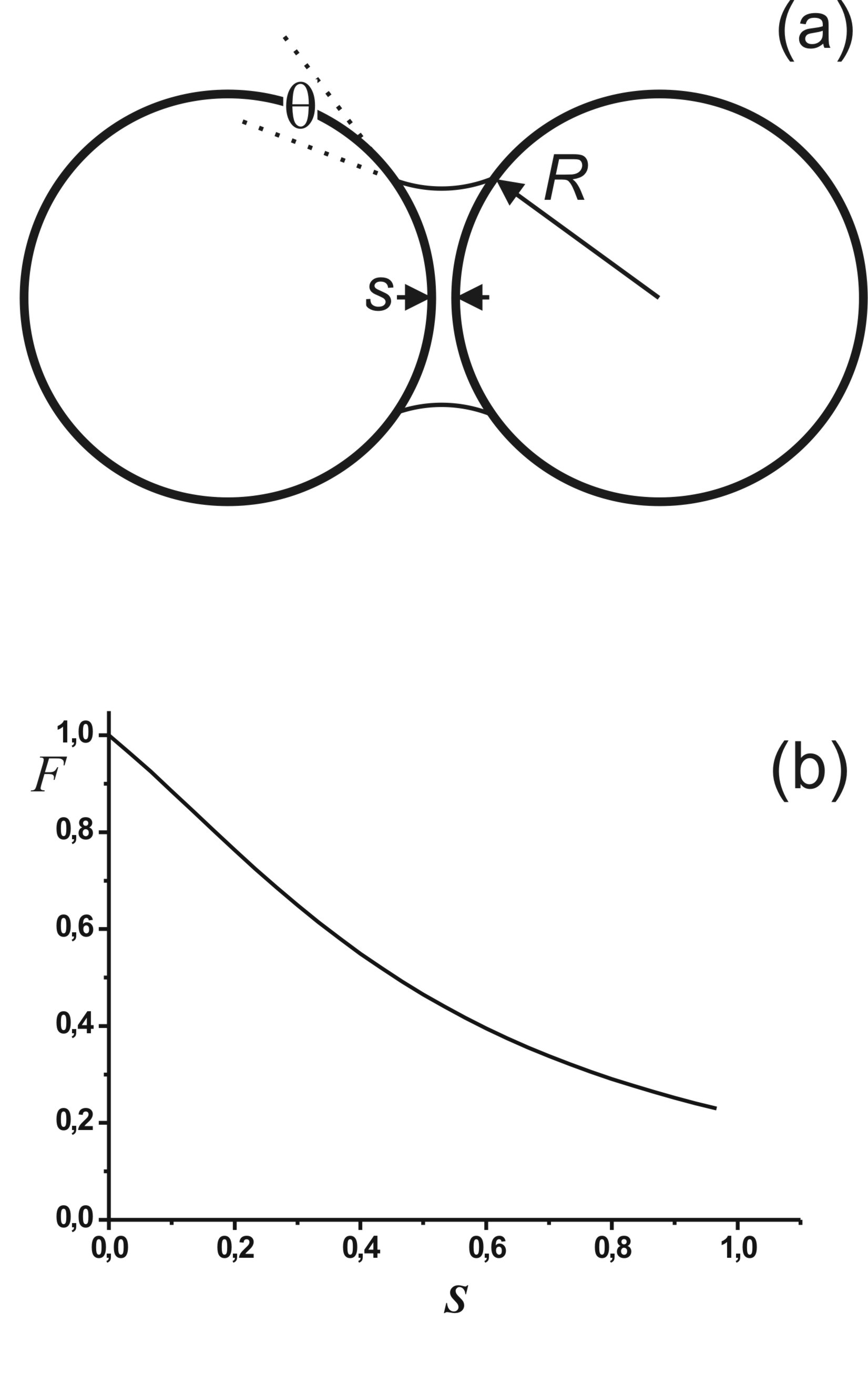}
\caption{(a) A liquid capillary bridge between two spherical
'grains' of radius $R$. The capillary bridge can span a certain
distance between the grains, until it pinches off at a critical
distance which depends upon the liquid volume in the bridge. (b) A
typical force-vs-distance curve observed a liquid bridge has formed.
It is shown as normalized with respect to the contact force, $2\pi
\gamma R \cos\theta$. \label{BridgeFigure} }
\end{figure}

On the contrary, the liquid capillary bridges which are present in a
wet granulate do provide their own energy scale. This is due to
their characteristic dynamics of bridge formation and rupture. When
two wet grains approach each other, the liquid adsorbed on their
surface will not react until they come into contact. At this point,
liquid is rapidly dragged to the area of contact due to the
interfacial forces, and a capillary bridge forms. When the grains
withdraw from each other after the impact, the bridge remains intact
for quite some distance, exerting an attractive force upon the
grains. This is illustrated in Fig.~\ref{BridgeFigure}a for the
idealized case of spherical grains. The angle $\theta$ is the
contact angle the liquid makes with the grain material, and
characterizes its wetting properties. For complete wetting, we have
$\theta = 0$. At a certain critical separation of the grain
surfaces, which depends upon the liquid volume of the capillary
bridge, the latter ruptures and distributes its liquid content back
onto the grain surfaces. If $\tilde{V} = V/R^3$ is the normalized
liquid volume of a bridge between spherical grains of radius $R$,
and $\tilde{s} = s/R$ is the normalized separation of the grain
surfaces, rupture occurs at
\begin{equation}
\tilde{s}_c = (1 + \frac{\theta}{2}) \left( \tilde{V}^{1/3} +
0.1\tilde{V}^{2/3} \right) \label{CritSep}
\end{equation}
in good approximation \cite{Willett00}.

A typical force-vs-distance curve is shown schematically in
Fig.~\ref{BridgeFigure}b. The force is normalized with respect to
the contact force, $F_0 = 2\pi R \gamma \cos \theta$, where $\gamma$
is the surface tension of the liquid \cite{Willett00}. Since there
is no liquid bridge (and thus no force) when the grains are
approaching, the energy lost in the entire process of formation and
rupture of the liquid bridge is given by the area under the
descending curve,
\begin{equation}
\Delta E_{cap} = \int_0^{s_c} F(s) \ ds \label{DEcapInt}
\end{equation}

For spherical grains, the shape of this curve is well known in the
quasi-static case \cite{Willett00}. It corresponds to the force
exerted by a rotationally symmetric minimal surface spanned between
the spheres, where the liquid volume of the bridge and the contact
angle are the main geometric parameters. A good approximation is
\begin{equation}
F = \frac{F_0}{1 + 1.05 S + 2.5 S^2}
\label{ForceDistance}
\end{equation}
where $S = \tilde{s}/\sqrt{\tilde{V}}$ \cite{Willett00}. This is in
fact the curve displayed in Fig.~\ref{BridgeFigure}b. It terminates
at $S_c = \tilde{s}_c/\sqrt{\tilde{V}}$ as given by
eq.~(\ref{CritSep}). Using this approximation, we can evaluate the
integral and obtain
\begin{equation}
\Delta E_{cap} = 0.67 \ F_0 R \sqrt{\tilde{V}}\ \arctan\left( 0.35 +
1.68 S_c \right)
\end{equation}
for complete wetting ($\theta = 0$). $S_c$ is at least of order
unity, such that $\arctan (...) \approx 1.1$ or larger, and it never
goes beyond 1.57. Hence a reasonable approximation is
\begin{equation}
\Delta E_{cap} \approx F_0 R \sqrt{\tilde{V}} \label{DEcap}
\end{equation}
Given the fact that we have neglected side effects such as contact
angle hysteresis, this should be as good as it gets.

 The presence of
a defined energy loss, which does not scale with the impact energies
of the colliding grains, has dramatic consequences for the
collective physical properties of the system. Most prominently, it
leads to phase transitions which occur when the granular temperature
is comparable to the energy loss
\cite{Kudrolli00,Geromichalos03,Herminghaus05,Fingerle07}. It is
illustrative to express the fixed energy loss in terms of an
energy-dependent restitution coefficient. We readily obtain
\begin{equation}
\varepsilon_{cap} = \sqrt{1-\frac{\Delta E_{cap}}{E}}
\label{epsiloncap}
\end{equation}
where $\Delta E_{cap}$ is constant. Obviously, $\varepsilon$ becomes
zero when $E = \Delta E_{cap}$. As a consequence, an energy-driven
phase transition occurs when the granular temperature comes close to
$\Delta E_{cap}$ \cite{Fingerle07}. This is not the case in a dry
system described by a constant restitution coefficient.

If the impact dynamics is sufficiently slow as compared to the
dynamics of bridge formation and rupture, we may assume that the
dynamic force-vs-distance curve corresponds to the quasi-static
case, as represented by eq.~(\ref{ForceDistance}). However, when we
consider dynamical processes, as they take place in a sheared or
otherwise agitated wet granular material, we have to discuss the
influence of this dynamics on $\Delta E_{cap}$.

In doing so, we will also have to consider the energy loss due to
viscous damping in the liquid. In reasonable approximation, the
viscous force is given by \cite{Ennis91}
\begin{equation}
F_{visc} = \frac{3 \pi}{2}R^2\eta \frac{v}{s}
\end{equation}
for spherical grains, where $\eta$ is the viscosity of the liquid
and $v$ is the relative velocity of the grains at impact. If viscous
forces are dominant, the equation of motion during withdrawal reads
$F_{visc} = -m \ddot{s}$, with the grain mass $m$, and the dot
indicating the derivative with respect to time. Direct integration
leads to
\begin{equation}
\Delta E_{visc} = m v_0 \int_{\delta}^{s_c}\frac{\dot{s}}{s}ds =
\frac{1}{2}m v_0^2 \ln \frac{s_c}{\delta} \left( 2\frac{v}{v_0}-\ln
\frac{s_c}{\delta}\right) \label{DEvisc}
\end{equation}
where $v_0$ denotes the velocity directly after the impact and
$\delta$ is a cutoff parameter, which may be identified with the
roughness of the grains \cite{Herminghaus05}. $v_0 = 9\eta / 8 \rho
R$ is a characteristic velocity, where $\rho$ is the density of the
grain material. If $v = v_0 \ln \frac{s_c}{\delta}$, the grains
stick together in the sense that their kinetic energy is not
sufficient to supply the {\it viscous} energy required for reaching
the rupture distance, $s_c$, even at zero surface tension. If $v$ is
considerably larger, one obtains the known result
\cite{Herminghaus05}
\begin{equation}
\Delta E_{visc} \approx \frac{3 \pi}{2}R^2\eta v \ln
\frac{s_c}{\delta} \label{DEviscKnown}
\end{equation}
Eq.~(\ref{DEvisc}) may as well be expressed in terms of an
energy-dependent restitution coefficient as
\begin{equation}
\varepsilon_{visc} = 1 - \sqrt{\frac{E_0}{E}} \label{epsilonvisc}
\end{equation}
where $E_0 = \frac{m}{2}[v_0 \ln(s_c/\delta)]^2$ is the impact
energy below which sticking occurs. We see from
eqs.~(\ref{epsiloncap}) and (\ref{epsilonvisc}) that there are great
similarities between the capillary and the viscous effects, and both
strongly differ from the scale-free energy loss encountered with the
dry systems.

In order to investigate to what extent the concepts discussed above
apply under realistic conditions, we have performed an experiment
particularly designed to map the conditions in agitated wet granular
matter as closely as possible. As model grains, we have used
spherical glass beads with $R = 1$ mm. These fell freely in a closed
box containing humid air onto a wet glass plate. Initial heights
were about 2 cm, such that the impact velocities were on the order
of a few cm/sec, which corresponds to typical granular temperatures
in agitated granulates. The motion of the glass spheres was recorded
with a fast CCD camera, the images were subsequently analyzed by
standard image processing techniques.

Since all arguments put forward above concerning forces between two
spherical grains apply as well to forces between a sphere and a flat
wall, we have studied the latter because of its better experimental
accessibility. We just have to keep in mind that in formulas
developed for two spherical grains, the radius of the sphere must be
multiplied by two. This corresponds to the well-known Derjaguin
approximation, $R_{eff}^{-1} = (R_1^{-1}+R_2^{-1})/2$. Although this
is not an accurate expression for the system under study
\cite{Willett00}, it provides a reasonable approximation since other
effects, like contact angle hysteresis, give rise to larger
uncertainties \cite{Herminghaus05}.

\begin{figure}
\includegraphics[width = 9 cm]{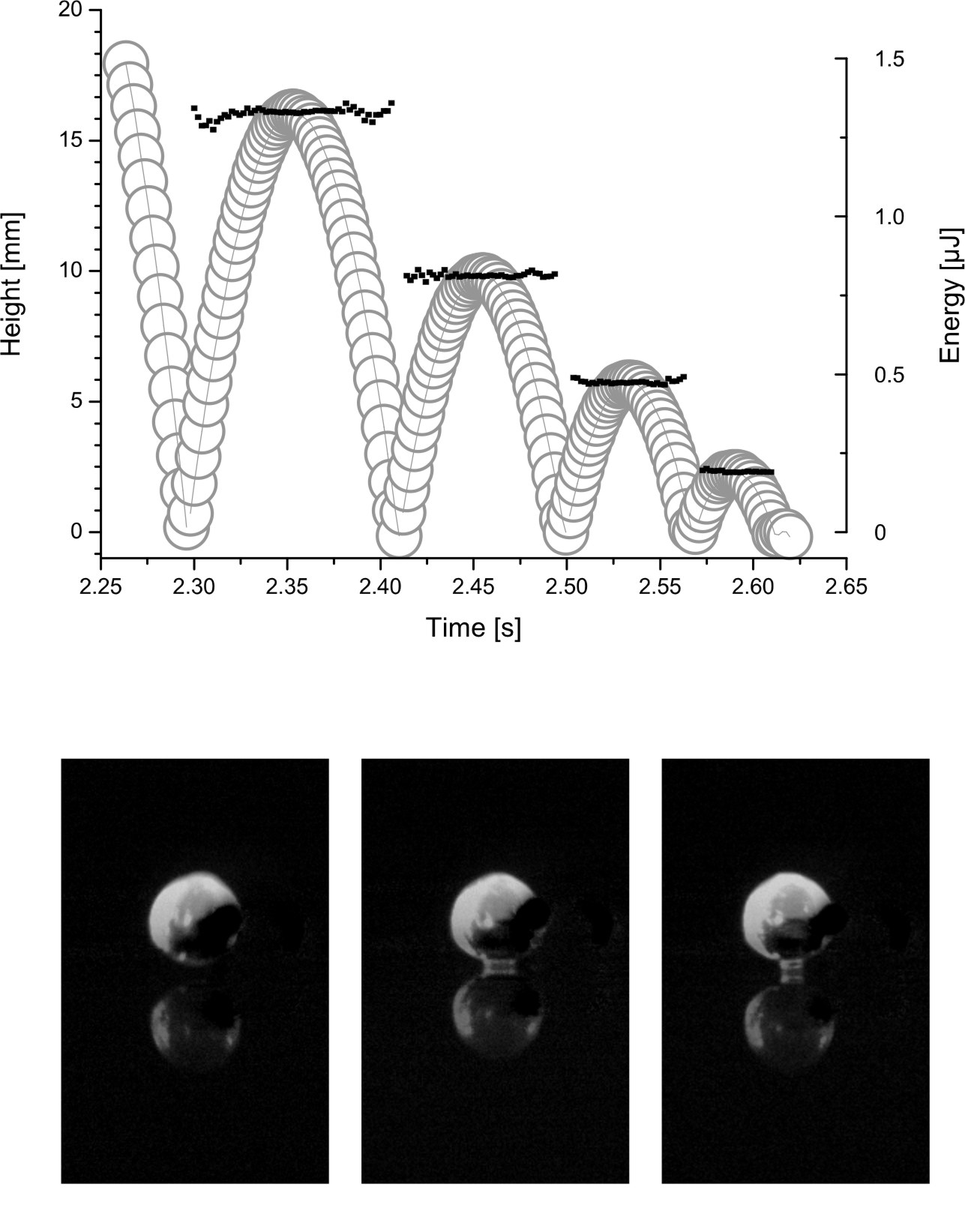}
\caption{Top: An experimental run as captured by a fast CCD camera
(491.3 frames/sec). From the trajectories, we can derive the total
energy as a function of time, which is shown as the small black
squares. Bottom: three consecutive images taken around the time of
impact with the bottom glass plate. The hysteretic character of the
bridge is clearly seen. The time elapsed between two consecutive
images is 2.04 ms. \label{Experiment} }
\end{figure}

The result is shown in Fig.~\ref{Experiment}. The height of a
bouncing glass sphere is plotted as a function of time (circles). On
the same time scale, the total energy as obtained from the
instantaneous velocity and height is indicated by the black squares.
While there is some scattering in the vicinity of the impacts,
mostly due to the finite delay between consecutive images, the
energy is observed to be constant with high accuracy away from the
impacts, allowing for an accurate determination of the energy level
of each bounce.

Below the main panel, we show three consecutive closeup images in
the ultimate temporal vicinity of an impact with the glass plate.
The time elapsed between the images is 2.04 ms. The hysteretic
character of the capillary bridge is clearly visible from its
absence before the impact (left) and its persistence afterwards
(right). From the difference in the energy levels of consecutive
bounces, we can deduce the energy lost in the impact with the glass
plate. For a quantitative analysis, we have to consider as well the
energy loss due to viscous friction in the air, $\Delta E_{air}$.
Using Stokes' formula, it is straightforward to see that
\begin{equation}
\Delta E_{air} \approx \frac{4 \pi R \bar{\eta}}{g} \left(
\frac{2E}{m} \right)^{3/2} \label{Stokes}
\end{equation}
where $\bar{\eta}$ is the viscosity of air. This is valid at low
Reynolds numbers. We have $Re \approx \sqrt{v/v_0} \approx
\sqrt{H/15.6 \mu m}$, where $H$ is the height of the bounce, such
that $Re$ reaches values around 30 in our experiment. Since
turbulence sets in only at much higher $Re$ \cite{Orszag00}, we can
safely assume eq.~(\ref{Stokes}) to describe our system well.

\begin{figure}
\includegraphics[width = 9cm]{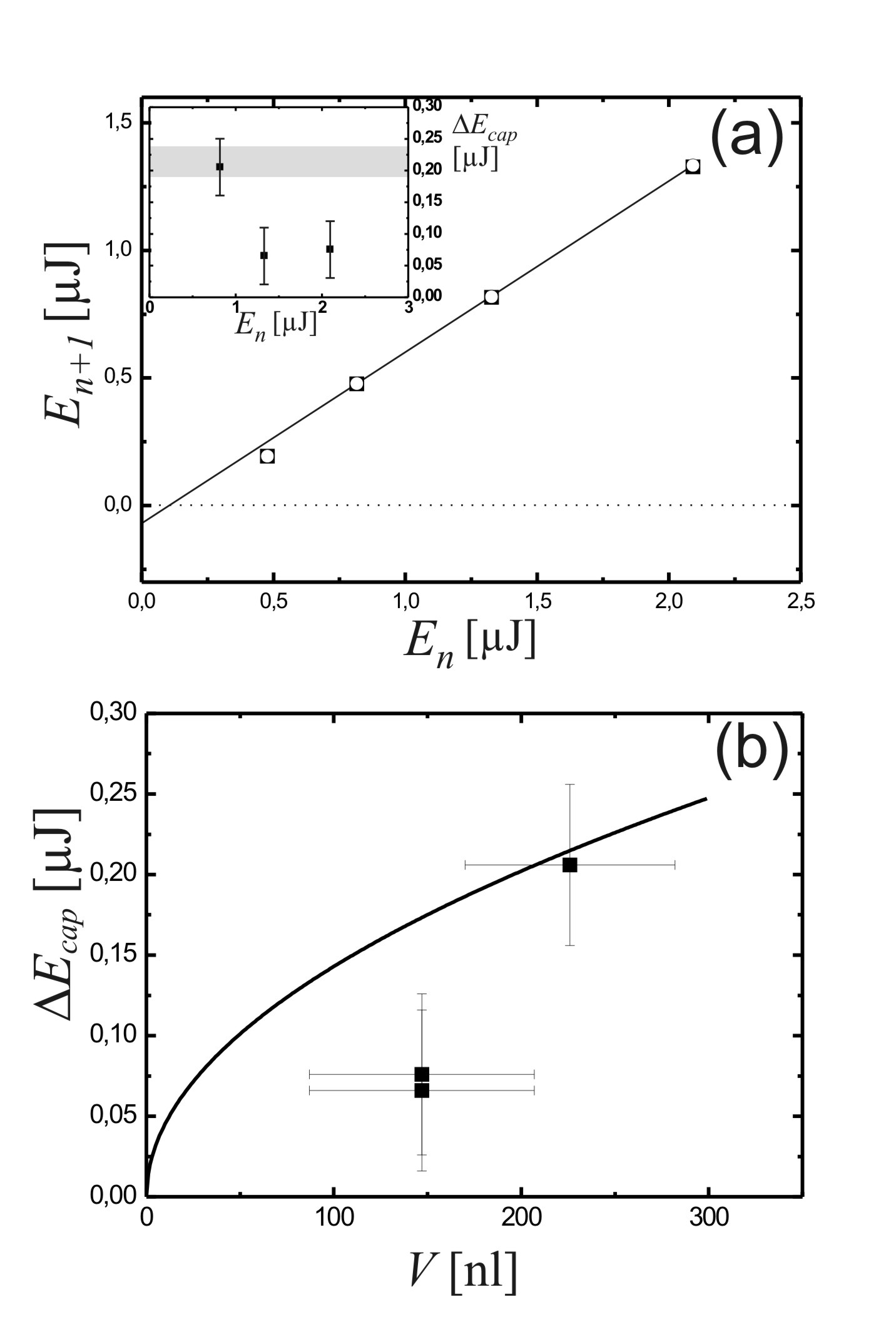}
\caption{(a) A 'return map' of the total energies between the
impacts. The accuracy of the measurement is sufficient for the
deviation of the datum point at smallest energy from the line to be
significant. It points to finite-time effects in the dynamics of
bridge formation and rupture. (b) The energy loss associated with
bridge rupture as a function of bridge volume, which was determined
independently from images taken around the impacts. The solid curve
represents the quasi-static limit (eq.~(\ref{DEcap})), without
fitting parameters. Although this equation relies on considerable
approximations, the deviation of the data points at smaller volume
is clearly significant.\label{Result} }
\end{figure}

The most convenient way of analyzing the data is to plot the energy
of each bounce as a function of the energy of the previous one.
Taking all dissipation mechanisms into account, we obtain
\begin{equation}
E_{n+1} = \varepsilon^2 E_{n} - \Delta E_{cap} - \Delta E_{visc} -
\Delta E_{air}
\end{equation}
where $n$ numbers the bounces, and the last two terms depend upon
$E_n$. The result is displayed in Fig.~\ref{Result}a. The full
squares represent the results for $E_{n+1}$ as obtained form the
experiment, while the open circles have been corrected for $\Delta
E_{air}$ according to eq.~(\ref{Stokes}), which is known without any
free parameters. As one can clearly see, the correction is of minor
importance, as as suggested already by the energy data from the
trajectories in Fig.~\ref{Experiment}a. From
eq.~(\ref{DEviscKnown}), we see that the viscous energy loss in the
liquid is of order $m v_0 v$, which is readily checked to be below 3
nJ in our experiments. It is thus even smaller than the viscous
dissipation in the air, and will henceforth be neglected. The
experimental error of the results displayed in Fig.~\ref{Result}a is
well below the size of the symbols.

The three data points for the largest energies lie on a straight
line within errors. From the slope of this line, we obtain the
restitution coefficient connected to the solid impact of the glass
sphere with the bottom plate. The result is $\varepsilon = 0.82 \pm
0.02$. Interpreting the intercept with the vertical axis as the
capillary bridge energy, we obtain $\Delta E_{cap} = 0.07 \pm 0.02
\mu$J. However, this intercept can be determined in principle for
each pair of consecutive energy levels, $E_n$. The corresponding
results for $\Delta E_{cap}$ are shown in the inset of
Fig.~\ref{Result}a as a function of the incident energy. The
deviation of the datum point at lowest energy from the straight line
in the main panel transforms into a pronounced increase of $\Delta
E_{cap}$ at low energy.

>From the few times at which we could capture the formation of a
bridge with the camera, we could estimate the bridge volume assuming
an axisymmetric shape. For the bridge at the slowest impact, we
obtained $\tilde{V} = 0.25 \pm 0.06$ for the dimensionless volume.
Together with eq.~(\ref{DEcap}) this leads to $\Delta E_{cap} =
0.214 \pm 0.056 \mu$J, where we have assumed complete wetting
($\theta = 0$). This result is indicated by the shaded area in the
inset of Fig.~\ref{Result}a and compares very favorably with the
experimental value obtained at low impact energy.

The reduction of $\Delta E_{cap}$ at larger impact energies may be
understood when one considers the dynamics of formation of the
bridge. There is some time needed for the liquid to rearrange from
the thick wetting layer around the contact point into the liquid
capillary bridge structure. The entire process spans several time
scales, starting from the microsecond range at individual asperities
of the grain roughness \cite{Zitzler02} to ripening processes on the
scale of minutes \cite{Kohonen04}. Consequently, although the impact
duration of the glass sphere with the bottom is sufficient to form a
small bridge, the volume of the latter will be larger when it has
more time to form. The characteristic time scale of the early stage
of bridge formation can be estimated from the typical height of the
bridge at contact, which is $\tilde{h} = h/R \approx
\sqrt{\tilde{V}/2\pi} \approx 0.2$. Knowing that viscous damping is
of minor importance, we consider the dispersion relation of undamped
capillary waves, $\omega^2 = \gamma q^3/\rho_{liq}$. With $q \approx
1/h$ we obtain the time scale $\tau = 1/\omega \approx 3.3 \times
10^{-4}$ sec. It is straightforward to calculate the kinetic impact
energy $E_{\tau}$ at which the solid surfaces are closer than $h$
for a duration equal to $\tau$. We obtain
\begin{equation}
E_{\tau} = \frac{8\pi R^2 \gamma \rho}{3 \tilde{h}\rho_{liq}}
\approx 780 nJ
\end{equation}
which is of the same order as the transition seen in the inset of
Fig.~\ref{Result}a.

However, the dynamics not only of bridge formation, but also of
bridge rupture gives rise to variations in $\Delta E_{cap}$. If the
solid surfaces withdraw rapidly, the formation of the extended neck
will be impeded and the bridge is expected to pinch off at a
separation which is smaller than the 'quasi-static' $s_c$. This
reduces the upper limit of the integral in eq.~(\ref{DEcapInt}), and
thus the value of $\Delta E_{cap}$. By measuring the bridge volumes
for different impacts independently, we can distinguish these two
effects. This is shown in Fig.~\ref{Result}b, where $\Delta E_{cap}$
is plotted as a function of the bridge volume, as determined from
images close to the respective impacts. The solid line represents
eq.~(\ref{DEcap}) and has no fitting parameters. Good agreement is
found for the larger bridge volume, which corresponds to the
leftmost point in the inset of Fig.~\ref{Result}a. At smaller
volume, $\Delta E_{cap}$ is indeed reduced, but this reduction in
is much stronger than predicted by the solid line, which represents
the quasi-equilibrium shapes.

If we finally return to the conversion of the energy loss into an
energy-dependent restitution coefficient, we see that the variations
in $\Delta E_{cap}$ will cause $\varepsilon(E)$ to decrease even
stronger with decreasing impact energy than suggested by
eq.~(\ref{epsiloncap}). This stresses again the qualitative
differences to the dry systems, where $\varepsilon(E)$ tends to be
decreasing with {\it increasing} impact energy. When one looks at
the results in the main panel of Fig.~\ref{Result}a, one might not
anticipate that the small negative intercept with the vertical axis
should be of any importance for the collective behavior of many
spheres. Quite surprisingly, the dramatic mechanical differences
between dry and wet sand show that this is nevertheless the case.

{\bf Acknowledgements}

Inspiring discussions with J\"urgen Vollmer are gratefully
acknowledged.


\begin{references}

\bibitem{Goldhirsch96} K. Sela, I. Goldhirsch, {\it Phys. Fluids} {\bf
9} (1996) 856.


\bibitem{Coniglio04} A. Coniglio et al., {\it Physica A} {\bf 339}
(2004) 1.

\bibitem{BrillPosch} N. Brilliantov, T. Poeschel, {\it Kinetic Theory of Granular Gases}
(Oxford Univ. Press, Oxford, UK, 2004).



\bibitem{BehringerSperl06} R.P. Behringer, K.E. Daniels, T. S. Majmudar, and M. Sperl,
{\it Fluctuations, Correlations, and Transitions in Granular
Materials: Statistical Mechanics for a Non-Conventional System}, in
Proc. 9th Experimental Chaos Conference (Sao Jose dos Campos,
Brazil, 2006).

\bibitem{Majmudar07} T. S. Majmudar, M. Sperl, S. Luding, and R. P.
Behringer, {\it Phys. Rev. Lett.} {\bf 98} (2007) 058001.


\bibitem{Ruiz02} Y. Nahmad-Molinari and J. C. Ruiz-Suarez, {\it Phys. Rev. Lett.}
{\bf 89} (2002) 264302.

\bibitem{AsteConiglio04} T. Aste, A. Coniglio, {\it Europhys. Lett.} {\bf
67} (2004) 165.

\bibitem{Ruiz05} O. Carvente and J. C. Ruiz-Suarez, {\it Phys. Rev. Lett.}
{\bf 95} (2005) 018001.

\bibitem{Keys07} A. S. Keys et al., {\it Nature Physics} {\bf 3}
(2007) 260.

\bibitem{FeitosaMenon04} K. Feitosa and N. Menon, {\it Phys. Rev. Lett.} {\bf
92} (2004) 164301.

\bibitem{Aumaitre04} S. Aumaitre, J. Farango, S. Fauve, and S.
McNamara, {\it Eur. Phys. J. B} {\bf 42} (2004) 255.

\bibitem{Herminghaus05} S. Herminghaus, {\it Adv. Phys.} {\bf 54} (2005) 221.

\bibitem{SandWorldDe} see, for instance: \begin{verbatim}
http:\\www.sandworld.de \end{verbatim}

\bibitem{Hornbaker97} D. J. Hornbaker et al., {\it Nature} {\bf 387}
  (1997) 765.

\bibitem{Mikami98} T. Mikami, H. Kamiya, M. Horio, {\it Chem. Eng.
    Sci.} {\bf 53} (1998) 1927.

\bibitem{Bocquet04} L. Bocuet, F. Restagno, and E. Charlaix, {\it Europhys. Lett.} {\bf
14} (2004) 177.

\bibitem{Scheel04} M. Scheel, D. Geromichalos, S. Herminghaus, {\it J.
    Phys.: Condens. Matter} {\bf 16} (2004) S4213.

\bibitem{Richefeu06} V. Richefeu, M. S. El Youssoufi, and F. Radjai,
{\it Phys. Rev. E} {\bf 73} (2006) 051304.

\bibitem{Duran00} J. Duran, {\it Sands, Powders, and Grains} (Springer, New York 2000)

\bibitem{Thornton98} G. Lian, C. Thornton, M. J. Adams, {\it
Chem. Eng. Sci.} {\bf 53} (1998) 3381.

\bibitem{Simons00} S. J. R. Simons, R. J. Fairbrother, {\it Powder
    Technol.} {\bf 110} (2000) 44.

\bibitem{Willett00} Ch. D. Willet et al., {\it Langmuir} {16} (2000)
  9396.

\bibitem{Kudrolli00} A. Samadani, A. Kudrolli, {\it Phys. Rev. Lett.} {\bf 85} (2000) 5102.

\bibitem{Geromichalos03} D. Geromichalos, M. Kohonen, F. Mugele, and
S. Herminghaus, {\it Phys. Rev. Lett.} {\bf 90} (2003) 168702.

\bibitem{Fingerle07} A. Fingerle, K. Roeller, K. Huang, and
S. Herminghaus, submitted to {\it Phys. Rev. Lett.}
\bibitem{Ennis91} B. J. Ennis, G. Tardos, and R. Pfeffer, {\it Powder Technol.} {\bf 65} (1991) 257.

\bibitem{Orszag00} A. G. Tomboulides, S. A. Orszag, {\it J. Fluid
Mech.} {\bf 416} (2000) 45.


\bibitem{Zitzler02} L. Zitzler, S. Herminghaus, F. Mugele, {\it Phys. Rev. B} {\bf
66} (2002) 155436.

\bibitem{Kohonen04} M. M. Kohonen et al., {\it Physica A} {\bf 339}
(2004) 7.

\end{references}

\end{document}